# EFFICIENT CALL PATH DETECTION FOR ANDROID-OS SIZE OF HUGE SOURCE CODE

Koji Yamamoto[1] and Taka Matsutsuka[1]

[1] Fujitsu Laboratories Ltd., Kanagawa, Japan
{yamamoto.kouji,markn}@jp.fujitsu.com

***ABSTRACT***

*Today most developers utilize source code written by other parties. Because the code is modified frequently, the developers need to grasp the impact of the modification repeatedly. A call graph and especially its special type, a call path, help the developers comprehend the modification. Source code written by other parties, however, becomes too huge to be held in memory in the form of parsed data for a call graph or path. This paper offers a bidirectional search algorithm for a call graph of too huge amount of source code to store all parse results of the code in memory. It refers to a method definition in source code corresponding to the visited node in the call graph. The significant feature of the algorithm is the referenced information is used not in order to select a prioritized node to visit next but in order to select a node to postpone visiting. It reduces path extraction time by 8% for a case in which ordinary path search algorithms do not reduce the time.*

## 1. INTRODUCTION

Source code written by other parties, especially open source code, are often utilized to build developers' own software products and services. The developers merge other party's code as a library, or adds their own original functions into it in order to make their software high value with competitive development cost. In the latter case, it is the key to success to understand the overlook and details of the source code.

It is an effective approach of source code comprehension to recognize relationships between classes or methods in the code. The main kinds of relationships for imperative object oriented programming languages like Java and C++ are caller-callee relationship (as known as call graph [1]), data structure, and class inheritance. We believe grasping caller-callee relationship, especially a caller-callee relationship path (abbreviated "call path" hereafter), is one of the best entries to comprehend source code because it highlights outlook of behaviours of the executed code so as to emphasize which methods/classes have to be first investigated in detail. Call paths are acquired using static program analysis.

The size of open source code is increasing huge. For instance, open source version of Android OS source [2] consists of 50 to 100 million lines of code. In spite of that, the required time for static analysis of source code has been reduced drastically today. Understand ™ [3] by Scientific Toolworks, Inc., for example, consumes less than one tenth of static analysis time of the previous tools such as Doxygen [4] in our experience. Static analysis of huge scale source code written by other parties is now the realistic first step to comprehend them if the following problem is resolved.

Huge amount of analysis result of huge size source code is, however, still barrier to understand the source code for an ordinary development environment. A developer usually has a general type of laptop/desktop computer with tiny memory, at most 16GB. The memory does not store all the result if the target source code is for Android OS, 50-100 million lines of code or similar size of code. Actually a server with much more size of memory cannot treat the result efficiently. It takes much more time to extract a call path. It disturbs developers' source code comprehension.

Our contribution is a bidirectional search algorithm to extract a call path from a call graph of too huge source code to store all parse results of the code in memory. It reduces 8% of path extraction time for a case in which ordinary path search algorithms do not reduce the time. The first characteristic feature of the algorithm is it refers to a method definition in source code corresponding to the visited node in the call graph. The second significant feature is the referenced information is used not in order to select a prioritized node to visit next but so as to select a node to "postpone" visiting. They are dedicated to the search time reduction. In addition, the algorithm halves the required time for the aforementioned case if all the data is stored in memory, though it is far from a real situation.

In the rest of this paper, we explain call graphs themselves and graph search algorithms. After that, we introduce our bidirectional search algorithm for call graphs and its evaluations. Then we make some discussions followed by concluding remarks.

## 2. DEFINITIONS

### 2.1. Call Graph

A call graph is the directed graph where the nodes of the graph are the methods (in Java; the functions in C++) of the classes of the program; each edge represents one or more invocations of a method by another method [1]. The former method is referred to as a callee method, and the latter as a caller method. The staring node of the edge corresponds to the caller method. The ending node stands for the callee method.

An example of call graphs is depicted in Figure 1, which is retrieved from the source code shown in Figure 2. Each bold text part in Figure 2 corresponds to the node having the same text in Figure 1. Information on methods and their invocations is retrieved from source code using static analysis tools like as [3] and [4]. In practice graph could be more complicated: Each method might be invoked by other methods than method "transmit()"; Method "transmit()" itself might be invoked by some methods.

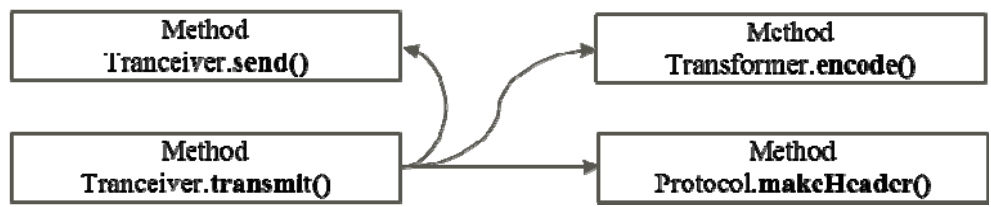

Figure 1. An example of call graphs

```
public class Tranceiver implements ITranceiver {
        private Transformer transformer;
        private Protocol protocol;
        private boolean send(Host destination) { ... }
        ...
        public boolean transmit(String data, Host destination) {
                byte[] encodedData = transformer.encode(data);
                header = protocol.makeHeader();
                send(header, encodedData, destination);
        }
}
```

Figure 2. Source code corresponding to the call graph in Figure 1

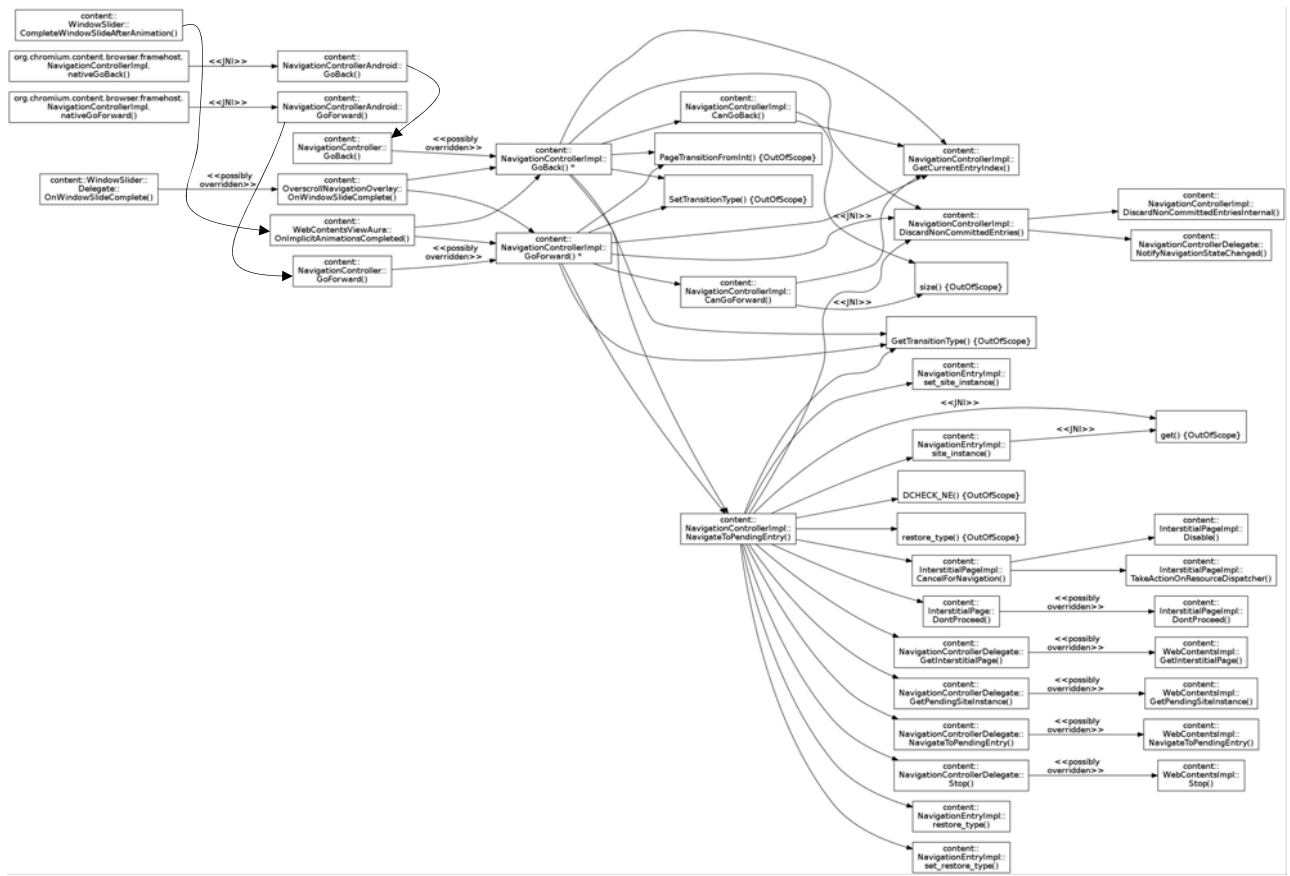

Figure 3. A complicated example of call graphs

## 2.2. Call Path

A call path is a sequence of edges where all the edges are connected and each edge is connected to at most one incoming edge and at most one outgoing edge. The edge having no incoming edge is called as an "initial node." The edge with no outgoing edge is called as a "final node." The methods corresponding to the initial node and the final node are called as an "initial method" and a "final method" respectively.

Figure 4 shows an example of call paths, which is extracted from a little bit complicated call graph shown in Figure 3. In most cases, extracted call paths are simpler to grasp caller-callee relationship than general call graphs for developers if they know the names of an initial method and a final method to be concerned.

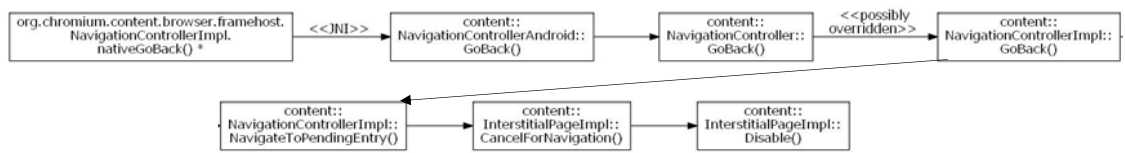

Figure 4. An example of call paths, which is extracted from the graph in Figure 3

## 2.3. Bidirectional Search Algorithm for Call Graphs

A bidirectional graph search algorithm is a graph search algorithm that finds a shortest path from an initial node to a final node in a directed graph. It traverses nodes forward in the graph and traverses nodes backward simultaneously [5]. Recent bidirectional algorithms use a heuristic distance estimate function to select a node to visit next [6] [7] [8]. An example of heuristic functions is Euclidean distance of a pair of nodes for the corresponding real distance is Manhattan distance.

To our knowledge, no heuristic estimate function for a call graph has been found yet.

## 3. BIDIRECTIONAL SEARCH ALGORITHMS FOR CALL GRAPH

We present an algorithm for bidirectional search in a call graph. The algorithm use source code properties corresponding to a visiting node in order to select a next node to visit, while other bidirectional search algorithms use a heuristic estimate value between a visiting node and the initial/final node [6] or frontier nodes [7][8]. In our algorithm 'bidir_postpone', a next visiting node is decided using the type of method corresponding to the visiting node in addition to the outgoing or incoming degree (outdegree or indegree) of the node.

```
procedure bidir_postpone(E, V, initialNode,
                         finalNode):
 1: todoF := {initialNode}
 2: todoB := {finalNode}
 3: /* Let prevF and prevB be dictionaries
      of type V to V.
      Let delay, distF, and distB be dictionaries
      of type V to integer. */
 4: for v in V do
 5:    prevF[v] := None; prevB[v] := None
 6:    delay[v] := 0
 7:    distF[v] := ∞; distB[v] := ∞ /* which
           stand for the distance from v along with
           forward (F)/backward (B) path. */
 8: endfor
 9: intermed := None
10: while |todoB| > 0 or |todoF| > 0 do
11:    if |todoB| < |todoF| then
12:       frwd := True; todo := todoF;
             frontiers := todoB; dist := distF
13:    else
14:       frwd := False; todo := todoB;
             frontiers := todoF; dist := distB
15:    endif
16:    todo2 := { }
17:    for node u in todo; do
18:       if delay[u] > 0 then
19:          delay[u] := delay[u] - 1
20:          add u to todo2
21:          continue for-loop with next node u
22:       else if the type of the class of
                   the method corresponding to u
                   is interface and not frwd then
                   not frwd then
23:          add u to todo2
24:          delay[u] := 3 - 1
25:          continue for-loop with next node u
26:       endif
27:       if frwd then
28:          next := {v | (u->v) in E}
29:       else
30:          next := {v | (v->u) in E}
31:       endif
32:       for node v in next; do
33:          alt := dist[u] + 1 /* all the edge
                   weighs one in a call graph. */
34:          if dist[v] > alt then
35:             prev[v] := u
36:             dist[v] := alt
37:             if v in frontiers then
38:                intermed := v
39:                exit from while
40:             endif
41:             add v to todo2
42:          endif
43:       endfor
44:       if frwd then todoF := todo2
45:       else todoB := todo2 endif
46:       todo2 := { }
47:    endfor
48: endwhile
49: if intermed is None then
50:    output ERROR
51: else
52:    v := intermid
53:    while prevF[v] is not None do
54:       output (prevF[v] -> v)
             as a path constituent
55:    endwhile
56:    v = prevB[intermid]
57:    while prevB[v] is not None do
58:       output (v -> prevB[v])
             as a path constituent
59:    endwhile
60: endif
```

Figure 5. Algorithm 'bidir_postpone'

Another difference between our algorithm and the previous algorithms is that the selected node by our algorithm is not a prioritized node to visit next but a node that is high cost to visit. The node will be scheduled to visit some steps later instead of visiting immediately.

Figure 5 shows our algorithm 'bidir_postpone.' E and V in parameters in the figure stand for a set of edges and a set of nodes in the call graph respectively.

At line 22, the algorithm determines next node to visit should be treated immediately or should be postponed treating. If the corresponding class type of the visiting node is interface in Java or abstract class in C++, the treatment is postponed. This type of method can be called by many methods. Therefore indegree of the corresponding node is greater than usual nodes. That is why visiting to such nodes should be postponed. If the visiting node is the case, treatment of the node will be suspended for 3 steps (See the line 24 and lines 18-20).

It is false that the postponement can be achieved by assigning heavy weight to edges adjacent to the nodes that hold the condition at line 22, instead of the treatment suspension because an algorithm using such heavy edges may output longer path than algorithm bidir_postpone.

## 4. EVALUATION

We compare the results of applying four types of algorithms to four pairs of initial and terminal nodes (hereupon the both nodes are referred to as starting point nodes) under two kinds of conditions. The result tells (1) if all the data is stored in memory, our algorithm reduces the duration for the significant case where the original bidirectional search does not reduce time compared to more naïve unidirectional search. (2) Even if the data is stored in HDD, our algorithm reduces the path extraction time by 8% for the aforementioned case. The details are shown hereafter.

### 4.1. Algorithms, data, and conditions to compare

The algorithms are the following four types. The second and third ones are almost the same as our algorithm itself:

A1 'Bidir_3postpone': It is the algorithm shown in Figure 5.

A2 'Bidir_6postpone': It is slightly modified version of algorithm of A1 'Bidir_3postpone' with delay 6. It delays node visiting for 6 steps at line 24 in Figure 5 instead of 3 steps. The purpose to compare the algorithm A1 with A2 is to check whether postponement step of algorithm A1 is adequate or not.

A3 'Bidir_0postpone': It is almost the same algorithm as A1 'Bidir_3postpone' and A2 'Bidir_6postpone.' In this algorithm, the condition at line 22 in Figure 5 is always false while the property in the source code, which is the type of the corresponding method, is retrieved from the parse result stored in HDD. The algorithm is for evaluation of an overhead of the parse result retrieval.

A4 'Bidir_balanced': It is the almost original version (appeared in [9]) of bidirectional search algorithm with no heuristic estimate functions, due to missing of estimate functions for call graphs. The difference between the algorithm and A1 in Figure 5 is that lines 18 through 26 are omitted and the rest of the for-loop starting from line 32 is always executed.

The starting point node pairs are as follows:

P1 'A->C': The number of reachable nodes by traversing forward from the initial node is much more than the number of reachable nodes by traversing backward from the final node.

P2 'C->N': The numbers of forward nodes from the initial node and backward nodes from the final node are both few.

P3 'N->R': Opposite pattern of P1 'A->C'. The number of forward nodes from the initial node is much less than backward nodes from the final node.

P4 'A->R': The numbers of forward nodes from the initial node and backward nodes from the final node are both many.

Figure 6 shows the numbers of nodes reachable from the initial node and the final node. For P2 and P4, the numbers of both type of nodes are almost the same. They are different from each other for P1 and P3. Note that the measurement of the number is logarithmic. For instance, the number for the final node for P3 is 5 times greater than one for the initial node.

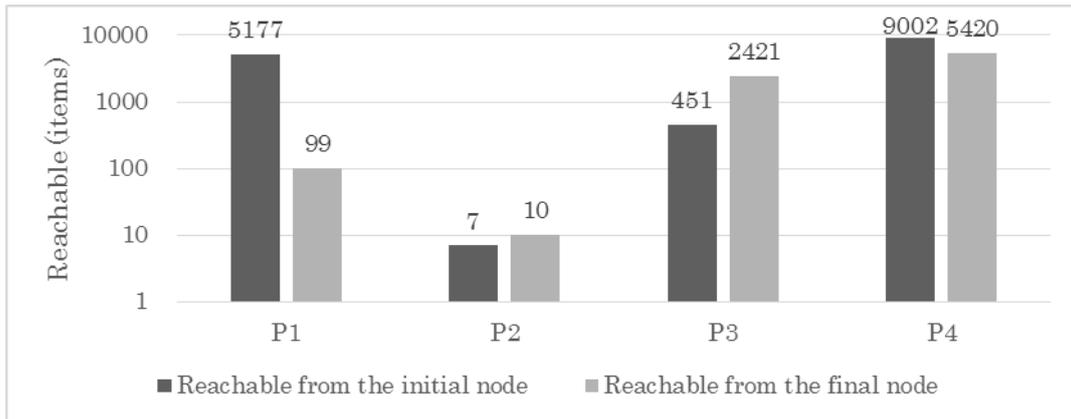

Figure 6. The numbers of nodes reachable from starting points

The conditions are the following two patterns. The latter is more similar to practical use case:

C1 'In memory': All graph data is stored in memory.

C2 'In HDD': A part of graph data is constructed on demand from the source code parse results that are stored in hard disk drive. Some graph edges and all graph nodes correspond to some parse result straightforwardly. Some other edges should be built up with syntactical analysis results and lexical analysis results.

C1 is the condition to measure the performance of the algorithms themselves. C2 is the condition that is almost the same as the actual execution environment with an exception. The environment contains parse result of source code stored in hard disk drive, and a result extraction program to convert specified part of the parse result to graph edges and nodes. The exception that differs from the actual environment is all the data stored in hard disk drive is not cached into memory (that is disk cache) at initial time. It makes measurement variance due to disk cache very small.

The source code for evaluation from which the call graph is constructed is partial source files of practical source code of Android OS for smartphones, version 4.4.4_r2 in [2]. The number of methods in the partial source files set is about 2% of the number in the whole source set. The parsed data for only 2% of the whole code occupies 2 GBytes size or more. For ordinary lap top computers of developers, even this size of partial source files are too huge to be held in memory.

### 4.2. Results

The measurements are executed 3 times. The average measured values are described in Figures 7 to 9, where logarithmic scale is used for all the Y axes. Figure 7 shows the time to traverse under the condition C1 'In memory'. Figure 8 is for the time to traverse under the condition C2 'In HDD'. Figure 9 tells the number of visited nodes. Note that an I-shaped mark at the top of each plotted box, in Figures 7 and 8, stands for the range of the sample standard deviation, +σ and -σ. Three times measurements seem enough because the deviations are sufficiently small.

Figure 7 depicts algorithm A1 'Bidir_3postpone' halves the time to traverse for the case in which the numbers of reachable nodes from the starting point nodes are both large (P4). The precise ratio is 1.07 seconds for our algorithm (A1) to 2.71 seconds for the original algorithm (A4), which stands for 60.5% reduction. Actually in other cases P1, P2, and P3, naïve bidirectional search (A4) runs in much shorter time than a unidirectional search, which is the directed edge version of Dijkstra algorithm, and is much more naïve than A4. Thus the time reduction in the case P4 is most desired by developers.

The resulting time to traverse for in-memory access case (C1, in Figure 7) is almost proportional to the number of nodes to be visited by each algorithm, shown in Figure 9. Therefore the less nodes are visited by the algorithm, the less traversal time can be achieved.

In contrast to the in-memory access, the results of the cases in which the data is stored in HDD (C2), in Figure 8, tell our algorithm takes worse time than the original algorithm (A4) for the case P3. In the case P4 that is most desired to reduce the time by developers, however, our algorithm (A1) spends 281.9 seconds and the original algorithm (A4) consumes 307.8 seconds. The difference is 25.9 seconds, which means ours (A1) achieves 8.4% reduction to the original (A4).

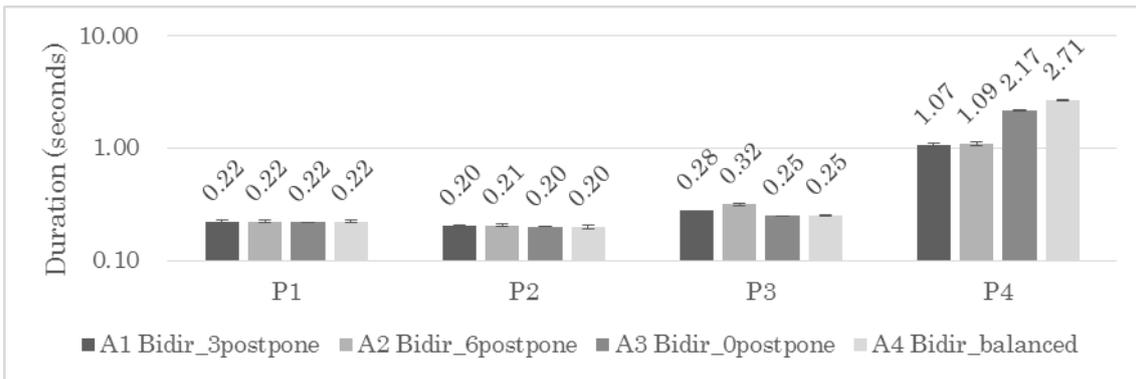

Figure 7. Time to traverse (in memory)

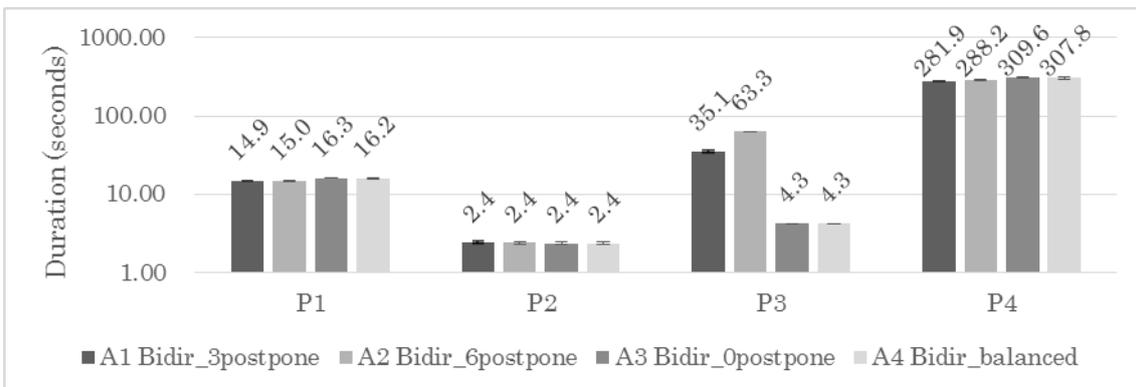

Figure 8. Time to traverse (in HDD)

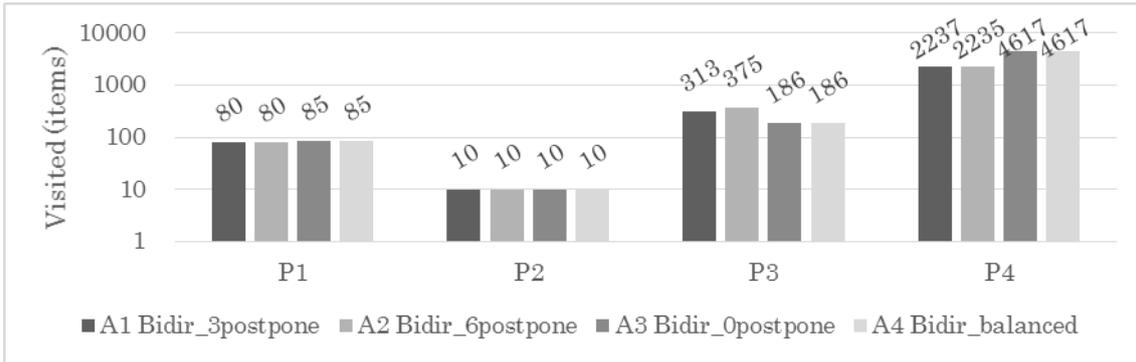

Figure 9. The numbers of visited nodes

### 4.3. Barriers to time reduction

We assume the overhead time for in-HDD case (C2, in Figure 8) comes from extra disk accesses to retrieve the properties of methods that occur at line 22 in Figure 5. Our disk access method in our algorithm is still naïve. Therefore the effect of the disk accesses could be reduced by the result of further investigation.

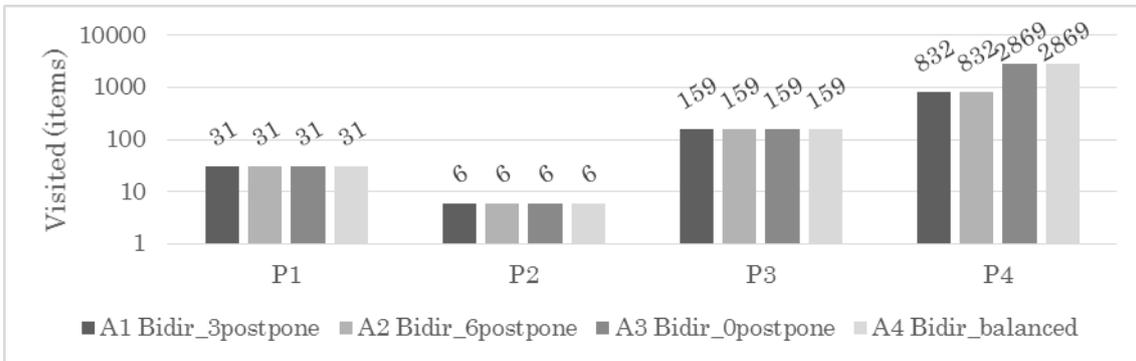

Figure 10. The numbers of nodes that the algorithms visited forward

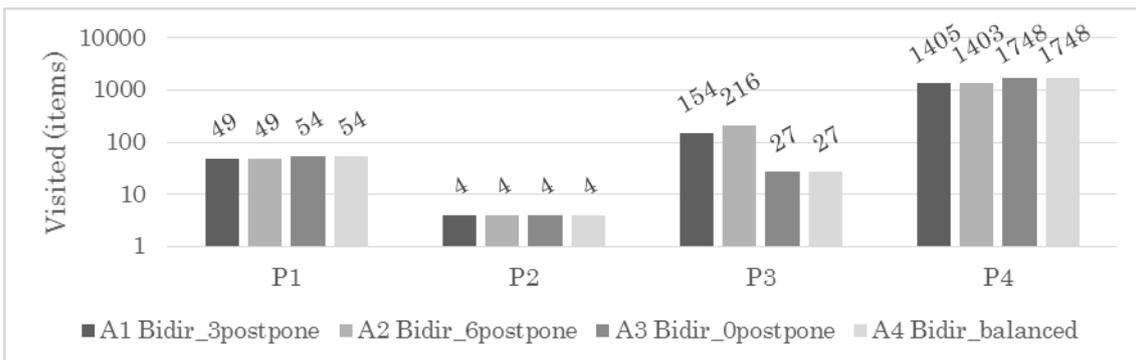

Figure 11. The numbers of nodes the the algorithms visited backward

The algorithm A1 visits remarkably large number of nodes for the starting point pair P3, in which the number of forward edges reachable from the initial node is much less than backward edges reachable from the final node. The algorithm traverses forward the same number of nodes from the initial node of P3 as the other algorithms (See Figure 10). It visits backward five times as large number of nodes from the final node of P3 as the algorithms A3 and A4, as shown in Figure 11. In the both algorithms A1 and A4, the forward search and the backward search meet at the same node. Note that the meeting point node is expressed as 'intermed' at line

38 in Figure 5. The node adjacent to the meeting point node (abbreviated as 'MP node' hereafter) in forward direction holds the postponement condition described at the line 22 in Figure 5. The adjacent node is on the path between the MP node and the final node. Thus visiting backward to the MP node is postponed for 3 steps. One hundred and twenty seven extra nodes has been visited while that. P3 notices our algorithm can be improved using other kinds of information than the type of the class corresponding to the visited node.

## 5. DISCUSSION

### 5.1. Intermediate Nodes

Actually resulting paths for our evaluation data P1 through P3 can be concatenated because the final node for Pn is the initial node for P(n+1) for n=1, 2. In addition, the concatenated path starts from the initial node of P4 and ends with the final node of P4.

It is interesting the summation of the numbers of visited nodes for P1 through P3 is much less than the numbers of visited nodes for P4 (See Figure 9.) It states the possibility of existence of somewhat "efficient" intermediate nodes for path search of source code call graph. If as bidirectional search algorithm can start traversing from the "efficient" intermediate nodes in addition to starting point nodes, it takes less time to extract call path than our current version of algorithm. We will try to find the feature of such type of intermediate nodes.

Note: The resulting path for P4, however, does not include the initial node and the final node of P2. Therefore we treat P1 through P4 as almost independent example data from each other.

### 5.2. More Aggressive Use of Features in Source Code

Our algorithm shown in Figure 5 only uses the class type corresponding to the visiting node as properties retrieved from source code. As discussed in the last section, there might be more types of properties that can reduce the number of nodes to be visited.

The result of further investigations on the kinds of useful properties of source code also affects the parsing way of source code itself. It will change the requirements of the source code parser. For example, some optional high cost feature of parsing will be executed by default. It will also change the database schema for the parse results to retrieve the useful properties with less cost.

## 6. RELATED WORKS

A bidirectional search algorithm in [6] or alike is called front-to-back algorithm. It uses the heuristic function to estimate the distance from the current visiting node (i.e., front) to the goal node (i.e., back). The function is similar to the heuristic function of A* search method [10]. The algorithm is executed under the assumption that the heuristic function estimates a value that is equal to or less than the real value (i.e., not overestimating.) The function is said to be admissible if the property holds.

A bidirectional search algorithm called front-to-front [7] [8] uses the heuristic estimate function that calculates the distance from the current visiting node (i.e., front) to the frontier node of the opposite direction search (i.e., (another) front.) The algorithm achieves the best performance when the function is admissible and consistent, that is, given nodes x, y, and z where x and z are the end nodes of a path and y is on the path, the estimated distance between x and z is equal to or less than the summation of the real distance between x and y and the estimated distance between y and z.

The former algorithm has been verified by experimental evaluations [6] for at least Fifteen-puzzle problems. The latter has been proved theoretically in [8].

For call graphs, however, either type of heuristic function has not been found yet to the best of our knowledge. Hence bidirectional search algorithms with heuristic estimate functions cannot apply to call graphs.

Although unfortunately we have not found previous works on call graph traversal especially related to bidirectional search, researchers of call graph visualizer made a comment on bidirectional search [11]. From experiences of participants attending to a lab study to evaluate the visualizer, they said "A significant barrier to static traversal were event listeners, implemented using the Observer Pattern. To determine which methods were actually called, participants would have to determine which classes implemented the interface and then begin new traversals from these methods." Our approach could resolve the difficulty and might make their traversal processes easier.

## 7. CONCLUSION

We have offered a bidirectional search algorithms for a call graph of too huge source code to store all parse results of the code in memory. It reduces 8% of path extraction time for a case in which ordinary path search algorithms do not reduce the time. The algorithm refers to a method definition in source code corresponding to the visited node in the call graph. The significant feature of the algorithm is the referred information is used not in order to select a prioritized node to visit next but in order to select a node to postpone visiting. They contribute to the search time reduction.

## ACKNOWLEDGEMENTS

We would like to thank all our colleagues for their help and the referees for their feedback.

**Authors**

Koji Yamamoto works at Fujitsu Laboratories, Japan, since 2000. He is interested in software engineering especially using program analysis, formal methods, and (semi-) automatic theorem proving and its systems. He received doctoral degree in engineering from Tokyo Institute of Technology, Japan, in 2000. He is a member of ACM.

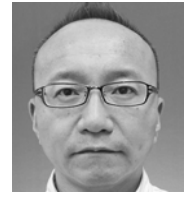

Taka Matsutsuka received his M.S. degree in Computer Science from Tokyo Institute of Technology. He works for Fujitsu Laboratories Ltd., and is engaged in R&D of OSS analysis. He is a visiting professor at Japan Advanced Institute of Science and Technology and a member of Information Processing Society of Japan.

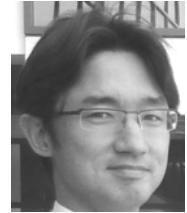